\documentclass[final,1p]{elsarticle}

\usepackage{lineno,hyperref}
\usepackage{amsmath,amsthm,amsfonts, amssymb}
\usepackage{algorithm}
\usepackage[noend]{algpseudocode}
\usepackage{subfigure}
\usepackage{color}

\makeatletter
\def\ps@pprintTitle{%
 \let\@oddhead\@empty
 \let\@evenhead\@empty
 \def\@oddfoot{}%
 \let\@evenfoot\@oddfoot}
\makeatother

\modulolinenumbers[5]

\newcommand{\vect}[1]{\mathbf{#1}}

\newcommand{\field}[1]{\mathbb{#1}}
\newcommand{\R}{\field{R}}

\begin{document}

\graphicspath{{figures/}}

\begin{frontmatter}

\title{Accelerated scale bridging with sparsely approximated Gaussian learning}

%\author{Ting Wang}
%\address{Aberdeen Proving Ground}

% or include affiliations in footnotes:
%\author[mymainaddress,mysecondaryaddress]{Ting Wang}

\author[add1]{Ting Wang}
\author[add1]{Kenneth W. Leiter}
\author[add2]{Petr Plech\'{a}\v{c}}
\author[add1]{Jaroslaw Knap}

\address[add1]{Simulation Sciences Branch, RDRL-CIH-C, U.S. Army Research Laboratory}
\address[add2]{Department of Mathematical Sciences, University of Delaware}
%%%%%%%%%%%%%%%%%%%%

\begin{abstract}
Multiscale modeling is a systematic approach to describe the behavior
of complex systems by coupling models from different scales. The
approach has been demonstrated to be very effective in areas of
science as diverse as materials science, climate modeling and
chemistry. However, routine use of multiscale simulations is often
hindered by the very high cost of individual at-scale
models. Approaches aiming to alleviate that cost by means of Gaussian
process regression based surrogate models have been proposed. Yet,
many of these surrogate models are expensive to construct, especially
when the number of data needed is large. In this article, we employ a
hierarchical sparse Cholesky decomposition to develop a sparse
Gaussian process regression method and apply the method to approximate
the equation of state of an energetic material in a multiscale model
of dynamic deformation. We demonstrate that the method provides a
substantial reduction both in computational cost and solution error as
compared with previous methods.
\end{abstract}


\end{frontmatter}

\section{Introduction}
\label{sec:introduction}
%%% Local Variables: ***
%%% mode:latex ***
%%% TeX-master: "acc_GP.tex"  ***
%%% End: ***

Multiscale modeling has now become a {\em de facto} standard approach
for the construction of high-fidelity models of complex phenomena and
systems encountered in many areas of science and
engineering~\cite{curtin2003Atomistic, pelupessy2013AMUSE,
  mahadevan2013High, suter2015Chemically, perdikaris2016Blood,
  alowayyed2016Multiscale}. The process of building a multiscale model
starts with identification of relevant phenomena occurring at
individual scales, both spatial and temporal. Thereafter, appropriate
at-scale models characterizing these phenomena are selected and
combined together into a single multiscale model. Computation is
fundamental to multiscale modeling as at-scale models are usually cast
in the form of computer models. In recent years, computational aspects
of multiscale modeling have become the focal point of numerous
research efforts (c.f.~\cite{groen18:_master} for an in-depth review
of recent developments). These efforts have led to a conclusion that
practicality of multiscale modeling hinges on the ability to
significantly reduce the often staggering computational cost of
at-scale models. Many different approaches have been proposed in order
to reduce this cost, with the vast majority falling under the name of
surrogate models. A surrogate model is a cheaper-to-evaluate
approximation of a model, constructed from direct observations of the
model.  Surrogate models have been employed with great success in
design optimization~\cite{wang2007review,koziel2011surrogate}, where a
model is repeatedly evaluated in the search for an optimal design. In
physical sciences, the use of surrogate models can be traced back to
the pioneering work of Pope~\cite{pope1997combustion}, who employed
surrogate modeling to enable simulations of combustion
chemistry. Other examples of the applications of surrogate models in
physical sciences include crystal
plasticity~\cite{barton2008Polycrystal, alharbi2015Crystal},
elastodynamics~\cite{rouet2014Spatial,roehm2015Kriging}, atomistic
modeling~\cite{li2015QM,leiter2018accelerated}, quantum
chemistry~\cite{rupp2012Fast, gomez2017Automatic}, and fluid
dynamics~\cite{zhao2018Active}. A comprehensive survey of surrogate
modeling techniques, including polynomial regression, kriging,
multivariate adaptive regression splines, polynomial stochastic
collocation, adaptive stochastic collocation, and radial basis
functions can be found in~\cite{jin2001comparative,
  sen2015evaluation}.

Gaussian process regression has been advocated as a particularly
flexible technique for surrogate model
development~\cite{knap2008adaptive,barton2011call,leiter2018accelerated}. However,
due to a significant cost of construction, Gaussian process regression
is rarely employed to build a single surrogate model. Instead, the
domain is often partitioned into a set of subdomains and separate
surrogate models are built over each of the subdomains. While such an
approach inevitably reduces the overall cost of constructing a
surrogate model, this reduction in cost may be accompanied by
considerable disadvantages, such as, for example, the loss of
smoothness of the surrogate model. In this article, we introduce a
methodology to reduce the cost of constructing surrogate models based
on Gaussian process regression and apply it in the context of
multiscale modeling.  We describe the multiscale modeling context of
our work in Section~\ref{sec:framework}. The details of our approach
are provided in Section~\ref{sec:accel-surr-model}, along with an
application of the technique to constructing a surrogate model of an
energetic material in Section~\ref{sec:comp-results}.

%%% Local Variables:
%%% mode: latex
%%% TeX-master: "acc_GP.tex"
%%% End:

\section{A computational framework for scale-bridging in multi-scale simulations}
\label{sec:framework}
%%% Local Variables: ***
%%% mode:latex ***
%%% TeX-master: "acc_GP.tex"  ***
%%% End: ***

\begin{figure}
  \centering
  \includegraphics[width=0.15\textheight]{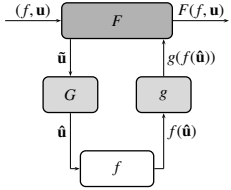}
  \caption{A two-scale model consisting of macroscale model $F$ and
    microscale model $f$. Two mappings transform data between scales:
    the input filter $G$ which transforms data into an appropriate
    form for the microscale model and the output filter $g$ which
    extracts relevant data from the microscale model to inform the
    macroscale model.}
  \label{fig:two_scale_model}
\end{figure}

The overarching context of the developments presented in this article
is the scale-bridging framework for multiscale modeling of Leiter~{\em
  et al.}~\cite{leiter2018accelerated}. Here, we only give a brief
description of the framework, the reader is referred
to~\cite{leiter2018accelerated} for a full exposition. The most
elemental multiscale model consists of two at-scale models, the
macroscale model $F$ and the microscale model $f$
(c.f. Figure~\ref{fig:two_scale_model}). The macroscale model is a
mapping $F:I \times D \mapsto R$, where $I$ is a collection of
microscale models, domain $D \subset \R^{H}$, and range $R \subset
\R^{\Xi}$. Similarly, the microscale model is a mapping $f:\hat{D} \mapsto
\hat{R}$ where $\hat{D} \subset \R^{\eta}$ and $\hat{R} \subset \R^{\xi}$
denote the domain and range of $f$, respectively. In addition, the
framework includes two mappings to transform data between at-scale
models. The mapping $G: \tilde{D} \mapsto \hat{D}$, where $\tilde{D}
\subset \R^{\tilde{\eta}}$ is the set of intermediate values derived from
values in $D$ by $F$. Henceforth, we refer to $G$ as the ``input
filter'' since it generates the input to $f$ in the set
$\hat{D}$. Likewise, the mapping $g: \hat{R} \mapsto \tilde{R}$, where
$\tilde{R} \subset \R^{\tilde{\xi}}$, is referred to as the ``output
filter'' as it extracts relevant data from the microscale model output
to be passed to the macroscale model. More complex multiscale models
can, of course, be formed through assemblies of multiple two-scale
model building blocks.

\begin{figure}
  \centering
  \subfigure[]{\includegraphics[width=0.15\textheight]{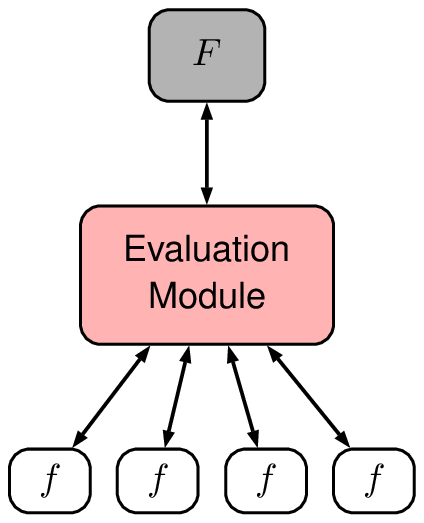}}
  \qquad
  \qquad
  \subfigure[]{\includegraphics[width=0.15\textheight]{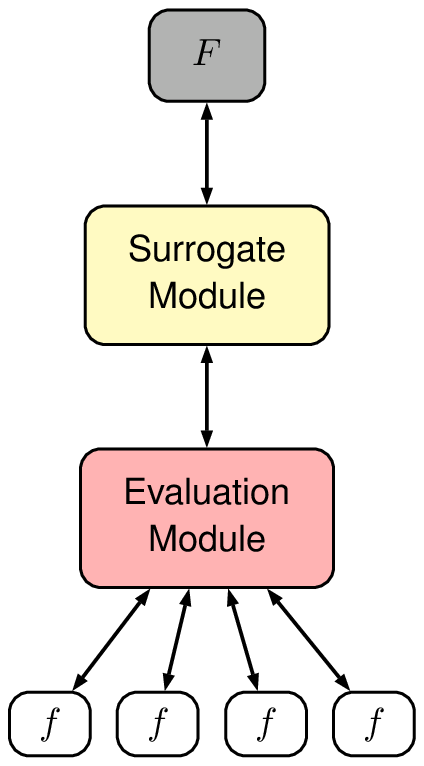}}
  \caption{A two-scale model with macroscale model $F$ and microscale
    model $f$. (a) The Evaluation Module is inserted as to facilitate
    evaluations of $f$ required by $F$. (b) The Surrogate Model is
    added to adaptively construct a surrogate model for $f$.}
  \label{fig:two_scale_model_surrogate}
\end{figure}

The centerpiece of the scale-bridging framework is a module
coordinating data exchanges between at-scale models, the Evaluation
Module (c.f. Figure~\ref{fig:two_scale_model_surrogate}~(a)). The act
of sending of $\vect{\tilde{u}} \in \tilde{D}$ from $F$ to the
Evaluation Module is denoted as an evaluation request. The Evaluation
Module carries out five distinct tasks: 1) it collects requests for
evaluation of $f$ from $F$; 2) it applies the input filter to the
evaluation requests to prepare input data for microscale models; 3) it
schedules evaluation requests on available resources; 4) it monitors
progress of evaluations to detect completion and handle failures; and
5) it applies the output filter to extract relevant data from
completed $f$ evaluations to return to $F$.  However, in many
practical applications, microscale models may be extremely costly to
evaluate and methods to lower the evaluation cost are necessary in
order to render the approach feasible. A popular approach, pioneered
by Pope~\cite{pope1997combustion} in combustion modeling, relies on
adaptive surrogate modeling, where evaluation requests are utilized to
on-the-fly build an approximation to the microscale model. Such an
approach is particularly advantageous as the modular structure of the
scale-bridging framework allows to incorporate surrogate models with
ease. Therefore, the framework can be simply augmented by the
Surrogate Module operating along side of the macroscale model and the
Evaluation Module
(c.f. Figure~\ref{fig:two_scale_model_surrogate}~(b)). The role of the
Surrogate Module is to automatically construct a surrogate model from
completed microscale model evaluation data and subsequently employ the
surrogate model in place of microscale model evaluations when
appropriate. As a consequence, the use of the surrogate model in the
evaluation of the microscale model is fully transparent from the
viewpoint of the macroscale model.

The literature dedicated to surrogate modeling is extensive and a
thorough survey of surrogate modeling approaches can be found
in~\cite{jin2001comparative,sen2015evaluation}. In principle, all
surrogate modeling techniques are directly applicable for construction
of a surrogate model in the Surrogate Module. However, the Surrogate
Model imposes two crucial constraints on the choice of surrogate
model. First, an error estimate at new evaluation requests must be
available so that the Surrogate Module can choose when to evaluate the
surrogate model or the underlying microscale model. Second, the
surrogate model must allow for the incorporation of new data acquired
from the evaluation of $f(\vect{\hat{u}})$ without excessive
computational cost. If the computational cost associated with updating
the surrogate model is high, the use of the surrogate model may not be
advantageous. A particular choice of surrogate modeling satisfying
both of the above constraints and advocated by Leiter~{\em et al.} is
Gaussian process regression~\cite{rasmussen2005gaussian}. However, due
to the fact that the cost of the Gaussian process regression is
dominated by the inversion of the covariance matrix, a single
surrogate model over the entire domain $\hat{D}$ of $f$ is not
constructed. Instead, a number of independent surrogate models with
finite support are constructed within $\hat{D}$. It bears emphasis
that the selection of training points for the construction of these
surrogate models is not carried out {\em a priori}, but instead
directly induced by $F$ itself. As a result, the set of all training
points within $\hat{D}$ is highly irregular and the set of all
surrogate models does not necessarily cover $\hat{D}$ in its
entirety. In addition, since individual surrogate models are entirely
independent, any notion of global smoothness is absent, leading to
complications under the circumstances when global smoothness of
approximations to $f$ is required~\cite{barton2011call}. Yet, despite
of these drawbacks, the above surrogate modeling scheme is still
capable of yielding remarkable savings in terms of the computational
cost, enabling truly extraordinary
simulations~\cite{leiter2018accelerated}.

%%% Local Variables:
%%% mode: latex
%%% TeX-master: "acc_GP.tex"
%%% End:

\section{Accelerated surrogate model}
\label{sec:accel-surr-model}
\begin{figure}
  \centering
  \includegraphics[width=0.05\textheight]{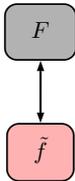}
  \caption{A simplified multiscale framework with the global surrogate
    model, $\tilde{f}$, interacting directly with the macroscale model
    $F$.}
  \label{fig:framwork_a_priori_surrogate}
\end{figure}

In this section, we describe an approach to substantially reduce the
cost of the construction of Gaussian surrogate models. We achieve this
goal by leveraging the hierarchical sparse Cholesky
decomposition recently developed by Sch{\"a}fer {\em et
  al.}~\cite{schafer2017compression} in order to build an approximated
global Gaussian surrogate model. Such an approach has two crucial
advantages over the approach of Leiter~{\em et
  al.}~\cite{leiter2018accelerated}. Namely, the global smoothness of
approximations to $f$ is guaranteed. Additionally, construction of
global surrogate models, i.e. over the entire $\hat{D}$, is
feasible. Hereafter, for simplicity, we focus on a two-scale model in
which the microscale model is replaced with an {\em a priori}
constructed surrogate model of the microscale model $\tilde{f}:\hat{D}
\mapsto \hat{R}$ (c.f. Figure.~\ref{fig:framwork_a_priori_surrogate}). We
emphasize, however, that the scenario considered here will likely not
always be applicable. For example, the microscale model may be
composed of two or more microscale models defined over subsets of
$\hat{D}$. Then, one may need to construct separate global surrogate
models over each of these subsets. Alternatively, one could construct
global surrogate models over some of these subsets and augment them
with surrogate models of the type considered in Knap~{\em et
  al.}~\cite{knap2008adaptive} and Leiter~{\em et
  al.}~\cite{leiter2018accelerated}. We
do not explore these scenarios in this article, but extensions of our
approach to them would be immediate.
For the clarity of presentation, we restrict the input dimension of the microscale model to be $\eta = 2$. 
However, the described approach in this section applies to microscale models with arbitrary input dimensions.

\subsection{Gaussian Process Regression} 
We briefly review the Gaussian process regression in this section.
Suppose we have a dataset $\mathcal{D} = (X, Y)$, where $X = (x_1, \ldots, x_N)^{\prime}$ represents a set of $N$ inputs in a domain $\Omega \subset \mathbb{R}^2$ and $Y = (y_1, \ldots, y_N)^{\prime}$ represents the corresponding outputs.   
We aim to build a regression model to interpolate the data set and then use 
the interpolant to predict at a new set of locations $X^*$. 
Gaussian process regression is a popular method for such regression problems~\cite{rasmussen2004gaussian}. 
We call a random process $\{f(x)\}_{x \in \Omega}$ a Gaussian process (GP) if
for any $n \in \mathbb{Z}_+$, the random vector $(f(x_1), \ldots, f(x_n))$ is a 
$n$-dimensional Gaussian random vector. 
The distribution of a GP is completely determined by its mean function $m(x)$ and its covariance function $k(x, x^{\prime})$ such that
\begin{equation}\label{eqn:Gaussian-mean}
m(x) = \mathbb{E}[f(x)]
\end{equation}
and 
\begin{equation}\label{eqn:Gaussian-cov}
k(x, x^{\prime}) = \mathbb{E}[(f(x) - m(x) ) (f(x^{\prime}) - m(x^{\prime}))].
\end{equation}
For simplicity, we assume the GP is centered, i.e., $m(x) = 0$. 
Hence, a centered GP $\{f(x)\}_{x \in \Omega}$ with covariance function $k(x, x^{\prime})$ can be written as
\[f(x) \sim \text{GP}(0, k(x, x^{\prime})).\]
We  denote $\mathcal{K}$ the covariance operator 
that acts on functions such that, for any suitable function $g$ (e.g., $g \in L^2(\Omega)$),
\[
(\mathcal{K}g) (x) = \int_{\Omega} k(x, x^{\prime}) g(x^{\prime}) \,dx^{\prime}.
\]
Often the measurement $y$ at location $x$ contains noise and hence we write
\[y(x) = f(x) + \epsilon,\]
where $\epsilon$ is assumed to be iid Gaussian noise $\mathcal{N}(0, \sigma_n^2)$ that is independent with the GP $f(x)$. 
Thus, the GP model with noisy measurement turns out to be
\[
y(x) \sim \text{GP}(0, k_{\sigma_n^2}(x, x^{\prime})),
\]
where $k_{\sigma_n^2}(x, x^{\prime}) =  k(x, x^{\prime}) + \sigma_n^2 \delta(x, x^{\prime})$ and $\delta(x, x^{\prime})$ is the Dirac function such that $\delta(x, x^{\prime}) = 1$ if
$x = x^{\prime}$ and $\delta(x, x^{\prime}) = 0$ otherwise.

Now given the data set $\mathcal{D}$, the Gaussian regression treats the prediction at $X^*$
as the mean of the conditional distribution $f(X^*) | \mathcal{D}$.
Since the joint distribution of $(Y, f(X^*))$ is a joint Gaussian random vector
\[\mathcal{N}\left(0, \begin{bmatrix} k_{\sigma_n^2}(X, X) & k(X, X^*)\\
k(X^*, X) & k(X^*, X^*)   \end{bmatrix}\right),
\]
it is straightforward to verity that $f(X^*) | \mathcal{D}$ is still a Gaussian vector 
with mean 
\begin{equation}\label{eqn:predictive-mean}
\mathbb{E}[f(X^*)| \mathcal{D}] = k(X^*, X) k_{\sigma_n^2}(X, X)^{-1}Y
\end{equation}
and covariance
\begin{equation}\label{eqn:predictive-var}
\text{Cov}[f(X^*)| \mathcal{D}] = k(X^*, X^*) - k(X^*, X)k_{\sigma_n^2}(X, X)^{-1}k(X, X^*).
\end{equation}
Finally, the mean serves as the prediction of the regression function at locations $X^*$ and the covariance provides a quantification of the prediction uncertainty.

As shown in the formula \eqref{eqn:predictive-mean} and \eqref{eqn:predictive-var},
GP regression requires the inversion of the dense covariance matrix $K \triangleq k_{\sigma_n^2}(X, X)$, which is often numerically unstable.
In practice, 
one often applies the Cholesky decomposition to $K$ such that $K = L L^T$, where $L$ is a lower triangular matrix. 
Nevertheless, for a dataset with $N$ observations, the computational complexity of Cholesky decomposition scales as $\mathcal{O}(N^3)$, which becomes computationally prohibitive when $N \gg 1$.
Unfortunately, this is the typical case in most multiscale problems where 
the construction of a high-fidelity surrogate model requires a large number of data. 
There exist rich literature on reducing the $\mathcal{O}(N^3)$ complexity by approximating the covariance matrix
in order to accelerate the Gaussian regression for a large dataset.
Most of these approximation methods can be roughly classified into two categories: (1) low rank approximation through subsampling \cite{rasmussen2004gaussian} 
and (2) sparse approximation using inducing variables \cite{quinonero2005unifying}.
Recently, Sch{\"a}fer {\em et al\@.} \cite{schafer2017compression} proposed a novel approximated Cholesky algorithm based on the gamblet transformation \cite{owhadi2017multigrid} which reduces the $\mathcal{O}(N^3)$ bottleneck down to near linear complexity.
Moreover, the upper bound of the approximation error can be shown to be exponentially small with respect to some pre-specified parameter. 
The algorithm, hereafter referred to as the hierarchical sparse Cholesky decomposition algorithm, requires 
the data points to be approximately equally spaced over the domain $\Omega$,
which may limit its application to a general dataset.
However, in the context of the multiscale bridging framework, data points from the microscale model can be sampled {\em a priori} at any location.
Hence, we have the flexibility to sample data points
over an uniform grid with equal spacing so that the 
hierarchical sparse Cholesky decomposition algorithm can be applied. 
In this paper, we aim to build a global Gaussian surrogate model based on the sparse Cholesky decomposition algorithm in order to accelerate the multiscale bridging. 

\subsection{Gamblets}
The theoretical foundation of the hierarchical sparse Cholesky decomposition algorithm relies on 
the exponential localization property of 
a set of multi-resolution basis functions called gamblets \cite{owhadi2017multigrid, owhadi2015bayesian}.
We briefly go over the definition of gamblets and its important properties in this section.

Given a centered Gaussian process $\{f(x)\}_{x \in \Omega}$ with kernel $k(x, x^{\prime})$,
we define $\mathcal{L}$ the precision/inversion operator of the covariance operator $\mathcal{K}$ such that, for any suitable function $g$, 
\begin{equation}\label{eqn:precision-operator}
\mathcal{L} \mathcal{K} g(x) = \mathcal{L} \int_{\Omega} k(x, x^{\prime}) g(x^{\prime})\, dx^{\prime} =  g(x).
\end{equation}
The key observation is that the Gaussian process $f(x)$ satisfies the following 
equation 
\begin{equation}\label{eqn:SPDE}
\mathcal{L}f(x) = \Delta(x),
\end{equation}
where $\Delta$ is a centered Gaussian process with covariance operator $\mathcal{L}$. 
Also, it can be verified that the kernel $k$ is the Green's function of \eqref{eqn:SPDE}, i.e., 
\[
\mathcal{L}k(x, x^{\prime}) = \delta(x, x^{\prime}).
\]  
Now given a data set $\mathcal{D} = (X, f(X))$, we define a set of basis functions called gamblets, 
\begin{equation}\label{eqn:gamblet}
\psi_i(x) = \mathbb{E}\left[f(x) | f(x_i) = 1 ~\text{and}~ f(x_j)   = 0, \forall j \neq i\right],
\end{equation}
for $i  =  1, \ldots, N$, 
i.e., it is the conditional expectation of $f(x)$ given the observation $f(x_j) = 0$ for all $j \neq i$ and $f(x_i) = 1$. 
It can be easily verified that the conditional expectation of 
$f(x)$ given $\mathcal{D}$ is a linear combination of the gamblets 
\begin{equation*}
\mathbb{E}[f(x) | \mathcal{D}] = \sum_{i=1}^N f(x_i)\psi_i(x). 
\end{equation*} 
This is saying that the best ``guess" of $f(x)$ given the dataset $\mathcal{D}$
is a linear combination of the observed features $f(x_1), \ldots, f(x_N)$ if all the gamblets are known. 
%According to \cite{schafer2017compression}, these basis functions $\psi_i$ are called gamblets 
%because they form a basis for betting on the value of the solution of \eqref{eqn:SPDE}
%given the data set $\mathcal{D}$.
The following two properties of gamblets are crucial for deriving the hierarchical sparse Cholesky algorithm. 
\begin{itemize}
\item {\bf Representation:}
Each gamblet function admits the following representation in terms of the covariance kernel $k$, 
\begin{equation}\label{eqn:gamblets-representation}
\psi_i(x) = \sum_{j=1}^N K_{ij}^{-1} \int_{\Omega} k(x, y) \delta(x_j, y) \,dy,
\end{equation} 
where $K^{-1}$ is the inverse of the covariance matrix $K = k(X, X)$.

\item {\bf Exponential localization:} 
The gamblet functions are exponentially decaying in the sense that 
\begin{equation}\label{eqn:exp-localized}
|\langle \mathcal{L}\psi_i, \psi_j \rangle| \leq C e^{-\beta d(x_i, x_j)}
\end{equation}
for some constants $C, \beta> 0$, where $\langle f, g \rangle = \int_{\Omega} f(x) g(x)\, dx$ is the $L^2$- inner product and $d(x, y)$ is some distance between $x$ and $y$.
This property provides the theoretical foundation to sparsely approximate the dense covariance matrix.
\end{itemize}

\subsection{Multi-resolution gamblets and block Cholesky decomposition}
\begin{figure}
  \centering
  \includegraphics[width=0.6\textheight]{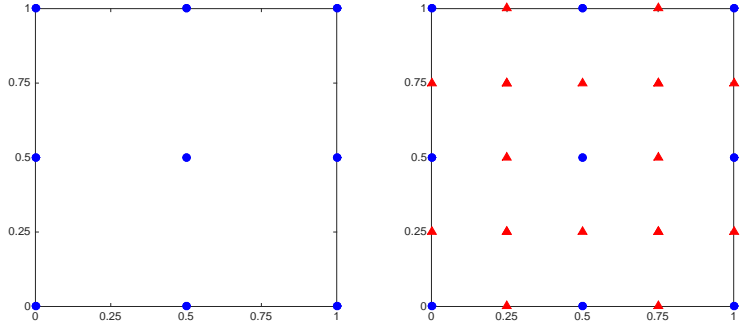}
  \caption{Configuration of dataset over $[0, 1] \times [0, 1]$. Left: the one-level uniform grid
  $X^{(1)}$ with resolution $2^{-1}$. Right: the two-level uniform grid $X^{(2)}$ with resolution $2^{-2}$.
  The index set $I^{(1)}$ contains the index of the blue dots and index set $I^{(2)}$ contains the index of both the blue dots and red triangles. 
 The index set $J^{(1)}$ contains the index of the blue dots and the index set $J^{(2)}$ contains the index of the red triangles.} 
  \label{fig:nested-grid}
\end{figure}

In this section, we provide a heuristic derivation for the hierarchical sparse Cholesky decomposition. 
The algorithm assumes that the configuration of the input set are nested with $q$ levels. 
For ease of presentation, throughout of this section we set $q = 2$ so that
the two-level input set $X = X^{(2)}$ forms an uniform grid of resolution 
$2^{-2}$ over the domain $\Omega = [0, 1] \times [0, 1]$. 
The two-level input set $X^{(2)}$ can be obtained by subdividing the level-one grid $X^{(1)}$, where $X^{(1)}$ is the set of input points that forms the uniform grid with resolution $2^{-1}$ (c.f.  Figure~\ref{fig:nested-grid}).
Similarly, the $q$-level uniform grid with resolution $2^{-q}$ can be obtained by recursively
subdividing $\Omega$ for $q$ times.
We denote $I^{(1)}$ and $I^{(2)}$ the index sets of points in $X^{(1)}$ and $X^{(2)}$ respectively.
That is, $x_i \in X^{(l)}$ whenever $i \in I^{(l)}$ for $l = 1, 2$.
We henceforth write $x_i^{(l)}$ to emphasize the fact that $x_i$ is a point in $X^{(l)}$.
Clearly, $I^{(1)}$ is a subset of $I^{(2)}$ and hence we can define the index set $J^{(1)} = I^{(1)}$ and $J^{(2)} = I^{(2)} / I^{(1)}$, i.e., $J^{(2)}$ contains the index of those data points that are in $X^{(2)}$ but not in $X^{(1)}$.
Hence, we can classify the points in $X^{(2)}$ into two categories: $X_{J^{(1)}}^{(2)} = X^{(1)}$ contains level two points that are also in level one and  $X_{J^{(2)}}^{(2)} = X^{2} / X^{(1)}$ contains points that are in level two but not in level one.

The above recursive sampling procedure can be viewed as a two-step hierarchical sampling approach, where a fine dataset $\mathcal{D}^{(2)} = (X^{(2)}, f(X^{(2)}))$ is sampled on top of the coarse dataset $\mathcal{D}^{(1)} = (X^{(1)}, f(X^{(1)}))$. 
Now given the datasets $\mathcal{D}^{(1)}$ and $\mathcal{D}^{(2)}$, 
we denote by
\[
K^{(1)} =k \left(    X^{(1)}, X^{(1)}   \right)
\] 
and 
\[
K^{(2)} = k \left(  X^{(2)}, X^{(2)}\right)
\] 
their corresponding covariance matrices.
Following the definition of gamblets in \eqref{eqn:gamblet},
we can define the level one and level two gamblets by 
\[
\psi_i^{(1)}(x) = \mathbb{E}\left[f(x) \left| f(x_i^{(1)}) = 1~\text{and}~f(x_j^{(1)}) = 0, \forall j \neq i \in I^{(1)}\right.\right]
\]
and
\[
\psi_i^{(2)}(x) = \mathbb{E}\left[f(x) \left| f(x_i^{(2)}) = 1~\text{and}~f(x_j^{(2)}) = 0, \forall j \neq i \in I^{(2)}\right.\right],
\]
respectively.
Using the definition of operator $\mathcal{L}$ and the representation \eqref{eqn:gamblets-representation}, it can be readily shown that the following two matrices
\[
B^{(1)} =  \left[\langle \mathcal{L}\psi_i^{(1)}, \psi_j^{(1)} \rangle  \right]_{i,j \in I^{(1)}}
\]
and
\[
B^{(2)} =  \left[\langle \mathcal{L}\psi_i^{(2)}, \psi_j^{(2)} \rangle \right]_{i,j \in I^{(2)}}.
\]
are the inverse of the covariance matrices
$K^{(1)}$ and $K^{(2)}$ respectively. 
Hence, hereafter we refer $B^{(l)}$ as the precision matrix associated with the dataset $\mathcal{D}^{(l)}$ for $l = 1, 2$.
It is immediate that each precision matrix is exponentially localized (i.e., nearly sparse) due to the exponential localization of gamblets.
Let us write the precision matrix $B^{(2)}$ in a block matrix form (corresponding to sets $X_{J^{(1)}}^{(2)}$ and $X_{J^{(2)}}^{(2)}$)
\[
B^{(2)}
=
\begin{bmatrix}
B_{11}^{(2)} & B_{12}^{(2)} \\
B_{21}^{(2)} & B_{22}^{(2)}
\end{bmatrix}.
\]
To be consistent, we also write the level one precision matrix $B^{(1)}$ as a single block matrix 
\[B^{(1)} = B_{11}^{(1)}.\] 
Owhadi and Scovel \cite{owhadi2017multigrid, owhadi2017universal} have proved that for a wide range of kernel functions $k$, 
the conditional numbers of the block matrices $B_{11}^{(1)}$ and $B_{22}^{(2)}$ are bounded. 
Based on this fact, they have further shown that the inverses matrices $B_{11}^{(1), -1}$ and $B_{22}^{(2), -1}$ are exponentially localized as well.

%\subsection{Exponentially localization of the Cholesky factor}
With the above preparations, now we are ready to motivate the hierarchical sparse Cholesky decomposition algorithm.
We start by making an important observation that links block Cholesky decomposition (or $LDL^T$ decomposition) with the two level hierarchical sampling procedure that we illustrated above.  
Recall that $K^{(2)}$ is the covariance matrix associated with the two-level dataset $\mathcal{D}^{(2)}$.
Its block Cholesky decomposition (corresponding to $X_{J^{(1)}}^{(2)}$ and $X_{J^{(2)}}^{(2)}$) reads
\begin{equation}\label{eqn:Cholesky-decomposition-1}
\begin{split}
K^{(2)} = 
\begin{bmatrix}
I & 0\\
K_{21}^{(2)}K_{11}^{(2), -1} & I
\end{bmatrix}
\begin{bmatrix}
K_{11}^{(2)} & 0\\
0 & K_{22}^{(2)} - K_{21}^{(2)}K_{11}^{(2), -1}K_{12}^{(2)} 
\end{bmatrix}
\begin{bmatrix}
I & K_{11}^{(2), -1}K_{12}^{(2)}\\
0 & I
\end{bmatrix},
\end{split}
\end{equation}
where $K_{i, j}^{(2)}$ is the covariance matrix between $X_{J^{(i)}}^{(2)}$ and $X_{J^{(j)}}^{(2)}$.
Since $K^{(2)}$ is the inverse of the precision matrix $B^{(2)}$, basic linear algebra shows that
\begin{equation}\label{eqn:block-mean}
K_{21}^{(2)} K_{11}^{(2), -1} = -B_{22}^{(2), -1} B_{21}^{(2)} = \left[\langle \delta(x_i^{(2)}, \cdot),  \psi_{j}^{(1)} \rangle  \right]_{i \in J^{(2)}, j \in J^{(1)}}
\end{equation}
and the Schur complement 
\begin{equation}\label{eqn:block-cov}
K_{22}^{(2)} - K_{21}^{(2)}K_{11}^{(2), -1}K_{12}^{(2)} = B_{22}^{(2), -1} = \left[\langle \mathcal{L}\psi_i^{(2)}, \psi_j^{(2)} \rangle \right]_{i, j \in J^{(2)}}.
\end{equation}
Hence the block Cholesky decomposition of $K^{(2)}$ can be rewritten in terms of $B^{(2)}$ as
 \begin{equation}\label{eqn:Cholesky-decomposition-2}
\begin{split}
K^{(2)} 
&= 
\begin{bmatrix}
I & 0\\
-B_{22}^{(2), -1} B_{21}^{(2)} & I
\end{bmatrix}
\begin{bmatrix}
B_{11}^{(1), -1} & 0\\
0 & B_{22}^{(2), -1}
\end{bmatrix}
\begin{bmatrix}
I & -B_{12}^{(2)}B_{22}^{(2), -1}\\
0 & I
\end{bmatrix}\\
&=
\begin{bmatrix}
L_{11}^{(1)} & 0\\
-B_{22}^{(2), -1} B_{21}^{(2)} & L_{22}^{(2)}
\end{bmatrix}
\begin{bmatrix}
L_{11}^{(1), T} & -B_{12}^{(2)}B_{22}^{(2), -1}\\
0 & L_{22}^{(2), T}
\end{bmatrix},
\end{split}
\end{equation}
where $L_{11}^{(1)}$ and $L_{22}^{(2)}$ are the Cholesky factors of $B_{11}^{(1), -1}$ and $B_{22}^{(2), -1}$ respectively, i.e., 
\[
B_{11}^{(1), -1} = L_{11}^{(1)}L_{11}^{(1), T}
\]
and 
\[
B_{22}^{(2), -1} = L_{22}^{(2)}L_{22}^{(2), T}.
\]
An application of Cauchy-Schwarz inequality to \eqref{eqn:block-mean} shows that 
the off-diagonal part of the Cholesky factor is exponentially localized. 
Furthermore, the result (Theorem~$3.12$) in \cite{schafer2017compression} shows that the block Cholesky factors $L_{11}^{(1)}$ and $L_{22}^{(2)}$ are also exponentially localized.
This suggests that the entire Cholesky factor in \eqref{eqn:Cholesky-decomposition-2} is exponentially localized.  
In other words, up to exponentially small entries, 
the Cholesky factor of $K^{(2)}$ is nearly sparse and the sparsity pattern is known a priori.
Therefore, it is desirable to skip the exponentially small entries in the process
of Cholesky factorization in order to reduce the computational complexity, which leads to the basic idea of the hierarchical sparse Cholesky decomposition.

\subsection{The sparse GP algorithm}
The above heuristic derivation is based on a dataset over a two-level uniform grid. However, the same argument can be easily extended for a 
dataset over a $q$ level uniform grid by successively doing the block Cholesky decomposition 
\eqref{eqn:Cholesky-decomposition-2} for $q-1$ times, which leads to the block Cholesky decomposition 
\[
K^{(q)} = L^{(q)}   L^{(q), T}.
\]
Since the locations (row and column) of those exponentially small entries in $L^{(q)}$ are explicitly known, with a certain confidence level, we can sparsely approximate $L^{(q)}$ by replacing these entries by zero. 
Sch{\"a}fer {\em et al.}~\cite{schafer2017compression} define the set of sparsity pattern to be
\[
\mathcal{S}_{R} = \left\{ (i, j) \in I^{(q)} \times I^{(q)} \left| i \in J^{(k)}, j \in J^{(l)}, |x_i - x_j| \leq R 2^{-k\wedge l}\right.\right\},
\] 
where $R$ is a parameter that controls the level of sparsity.
Then they suggest to restrict the Cholesky computation only to this sparsity set 
through applying the zero fill-in incomplete Cholesky decomposition \cite{golub2012matrix} to the sparse matrix
\begin{equation}\label{eqn:sparse-K}
K_{R}^{(q)}(i, j) = 
\left\{
\begin{array}{cc}
K^{(q)}(i, j) & \text{for}~(i, j) \in  \mathcal{S}_{R}\\
0 & \text{otherwise},
\end{array}
\right.
\end{equation}
such that $K_{R}^{(q)} \approx L_{R}^{(q)} L_{R}^{(q), T}$. 
Here the incomplete Cholesky factor $L_{R}^{(q)}$ is obtained by stepping through the Cholesky reduction on $K_{R}^{(q)}$ setting $L_{R}^{(q)}(i, j)$
to zero if the corresponding $K_{R}^{(q)}(i, j)$ is zero.
Finally, the main result in \citep{schafer2017compression} asserts that
we can use $L_{R}^{(q)} L_{R}^{(q), T}$ to approximate the original dense covariance
matrix $K^{(q)}$ such that
\begin{equation}\label{eqn:error-bounds}
\left\| K^{(q)} - L_{R}^{(q)} L_{R}^{(q), T} \right\| \leq p(N) e^{-\gamma R}
\end{equation}
for some constant $\gamma$ and polynomial $p(N)$, where the constant $\gamma$ is independent of $q$. 
Moreover, the computational complexity for obtaining $ L_{R}^{(q)}$ is 
$\mathcal{O}(N \log^2(N) R^4)$, which is near linear in $N$.
Hence, the sparsity parameter $R$ controls the trade-off between efficiency and accuracy. 
With larger $R$, the error bound is decreased but the complexity is increased. 
Finally, the sparse GP algorithm that uses the hierarchical sparse Cholesky decomposition for Gaussian regression is as follows. 
%%%%%%%% MAIN ALGORITHM %%%%%%%%
\begin{algorithm}
\caption{Sparse GP on an uniform grid of level $q$}\label{alg:main-algorithm}
\begin{algorithmic}[1]
\Procedure{SparseGP}{$\mathcal{D}^{(q)}$}
\State Order the data points from $J^{(1)}$ to $J^{(q)}$ (coarse to fine)
\State Initialize $K_{R}^{(q)}(i, j) = K^{(q)}(i, j)$ if $(i, j) \in S_{R}$ and $K_{R}^{(q)}(i, j)= 0$ otherwise
\State $L_{R}^{(q)}  \gets ~\text{IncompleteChol}(K_{R}^{(q)})$\Comment{Incomplete Cholesky decomposition}
\State $\alpha = L_{R}^{(q), T} \backslash L_{R}^{(q)} \backslash Y$\Comment{Backward substitution and then forward substitution}
\State $v = L_{R}^{(q)} \backslash k(X^{(q)}, X^*)$
\State $\mathbb{E}[f(X^*)| \mathcal{D}^{(q)}] = k(X^{(q)}, X^*)^T \alpha$\Comment{Update the predictive mean}
\State $\text{Cov}[f(X^*)| \mathcal{D}^{(q)}] = k(X^*, X^*) - v^T v$\Comment{Update the predictive variance}
\EndProcedure
\end{algorithmic}
\end{algorithm}
%%%%%%%%%%%%%%%%%%%%%%%%%%

A few comments are in order: 
$(1)$ Note that the above algorithm is simply the standard Gaussian regression with the Cholesky decomposition of $K^{(q)}$ replaced by the incomplete Cholesky of its sparse approximation $K_{R}^{(q)}$. 
Hence, the algorithm can be easily implemented on top of the standard GP algorithm.
$(2)$ It is important that the input points $X^{(q)}$ in the dataset are order from $J^{(1)}$ to $J^{(q)}$ for the algorithm to work correctly.
In the case that the data are not sample in this order, we can identity a permutation matrix $P$
such that $PKP^T$ has the correct ordering.   
$(3)$ Our presentation of Algorithm~\ref{alg:main-algorithm} assumes that the configuration of the input set is strictly uniform over the domain $\Omega$. 
However, we point out that this is requirement can be relaxed.  
Indeed, the algorithm is robust as long as the input configuration satisfies some certain criteria (c.f. \cite{schafer2017compression} for more details). 

\section{Results}
\label{sec:comp-results}
%%% Local Variables: ***
%%% mode:latex ***
%%% TeX-master: "acc_GP.tex"  ***
%%% End: ***

We now assess the sparse GP method in the context of a computationally
demanding multiscale model of impact physics. The multiscale model
consists of two at-scale models, a continuum mechanics macroscale
model and a particle-based microscale model. The surrogate model
serves to replace evaluation of the microscale model at a
significantly reduced computational cost. The new sparse GP method is
compared with the adaptive sampling based approach of Leiter~{\em et
al.}~\cite{leiter2018accelerated}, summarized below, which constructs
individual surrogate models on subsets of the entire training dataset.

\subsection{Macroscale model}

The macroscale model is a continuum mechanics model of a deforming
body implemented in the ALE3D multi-physics finite-element
code~\cite{anderson2003users}. The material of the body is taken to be
the energetic 1,3,5-trinitrohexahydro-s-triazine (RDX) and its
equation of state (EOS) is obtained through evaluation of the
microscale model. The EOS provides the pressure, $p$, and temperature,
$T$, for a given mass density, $\rho$, and internal energy density,
$e$.  The macroscale model uses a modified predictor-corrector
algorithm to integrate energy forward in time. The predictor step
includes the pressure volume work over the first half of the timestep
plus strain work from the deviatoric stress: $\tilde{e} = e_t -
\frac{1}{2}p_tdV + de_{dev}$, where $e_t$ is the energy density at the
start of the timestep, $p_t$ is the pressure at the start of the
timestep, $dV$ is the change in relative volume over the timestep, and
$de_{dev}$ is the change in deviatoric strain energy density. The
deviatoric strain energy is computed using a conventional
$J_2$-plasticity model with Steinberg-Guinan
hardening~\cite{steinberg1980} with a yield strength of 150 MPa,
hardening coefficient of 200 and hardening exponent of 0.1. The
predictor pressure and temperature are computed using the EOS as

\begin{gather}
  p_p = p(\rho_{t+1}, \tilde{e})
  \label{eq:non_reactive_EOS_pp} \\
  T_p = T(\rho_{t+1}, \tilde{e})
  \label{eq:non_reactive_EOS_Tp}
\end{gather}
where $\rho_{t+1}$ is the mass density at the end of the timestep.
The corrector step updates the energy as $e_{t+1} = \tilde{e} -
\frac{1}{2}p_pdV$ giving the pressure and temperature at the end of
the step as

\begin{gather}
  p_{t+1} = p(\rho_{t+1}, e_{t+1})
  \label{eq:non_reactive_EOS_p1} \\
  T_{t+1} = T(\rho_{t+1}, e_{t+1})
  \label{eq:non_reactive_EOS_T1}
\end{gather}

At modest pressures, the energy update in the corrector step is small,
which leads to a small change in pressure. In order to avoid a second
EOS evaluation per timestep, the pressure and temperature corrections
are omitted in our approach. The energy is updated, but the
temperature and pressure updates lag behind:

\begin{gather}
  p_{t+1} = p_p
  \label{eq:non_reactive_EOS_p_n+1} \\
  T_{t+1} = T_p
  \label{eq:non_reactive_EOS_T_n+1} \\
  e_{t+1}=\tilde{e}-\frac{1}{2}p_pdV
  \label{eq:non_reactive_EOS_e_n+1}
\end{gather}

\subsection{Microscale model}

The microscale model computes the EOS using energy-conserving
dissipative particle dynamics (DPD). The simulations are managed by
the LAMMPS Integrated Materials Engine (LIME), a Python wrapper to
LAMMPS that automates EOS evaluation~\cite{barnes2016lime}.  LIME
initializes and equilibrates a simulation cell containing 21,952
particles (28 x 28 x 14 unit cells) of RDX to be consistent with the
prescribed mass density and energy density. Following equilibration,
the temperature and pressure of the system are computed via ensemble
averages.

\subsection{Adaptive sampling}

The adaptive sampling method reduces computational cost of expensive
multiscale models. We refer the reader to Knap~{\em et
  al.}~\cite{knap2008adaptive} and Leiter~{\em et
  al.}~\cite{leiter2018accelerated} for a detailed description of the
adaptive sampling algorithm and its implementation in the
scale-bridging framework, but will briefly describe the method
here. Adaptive sampling is an active learning algorithm that
constructs a set of local GP surrogate models on-the-fly to replace
the evaluation of computationally expensive microscale models. In the
case of the multiscale model considered here, the surrogate models
approximate the EOS computed using DPD. While the macroscale model
integrates its solution forward in time, it repeatedly evaluates the
EOS. With adaptive sampling, the EOS is either evaluated by a
surrogate model or by the microscale model. An error estimate of the
surrogate model for a particular $\rho$ and $e$ is compared to a
user-specified acceptable error tolerance parameter,
$\tilde{e}_{tol}$, to determine if the EOS evaluation is satisfied by
the surrogate model. If the error estimate is too high, the EOS is
computed by the microscale model and the result is used to update the
surrogate model. The adaptive sampling algorithm continuously improves
the accuracy of the surrogate model for regions of the EOS of interest
to the macroscale model. The number of data to be incorporated into a
surrogate model may be potentially very large, especially for low
values of $\tilde{e}_{tol}$. The computational cost of GP regression
scales as $\mathcal{O}(N^3)$ where $N$ is the number of data. Rather
than construct a single GP surrogate model across all of the data, the
adaptive sampling algorithm builds a number of local GP surrogate
models on a partition of the overall data. A parameter $d_{max}$
determines the maximum number of data per local surrogate model. The
collection of surrogate models is stored in a metric-tree database to
allow quick access for evaluation and update.

Although adaptive sampling has been very successful in reducing
computational cost of expensive multiscale
models~\cite{knap2008adaptive, barton2008Polycrystal,
  leiter2018accelerated} it suffers two significant drawbacks in
practice: 1) the patchwork collection of local GP surrogate models
gives no guarantee of continuity between the patches; 2) the overall
simulation is often highly load imbalanced due to the unpredictable
adaptive execution of expensive microscale models.

\subsection{Taylor impact simulation}

We compare the performance of the sparse GP method and the adaptive
sampling method for the simulation of a Taylor impact experiment,
commonly used to characterize the deformation behavior of
materials~\cite{taylor1948}. The simulation setup is identical to the
one in~\cite{leiter2018accelerated}. In a Taylor impact experiment, a
cylinder of material travels at a constant initial velocity and
impacts a rigid anvil. In the simulations presented here, the cylinder
of RDX has a height of 1.27~cm and radius of 0.476~cm and travels at
200~m/s. In the macroscale model, axisymmetry is imposed along the
cylinder axis and the cylinder is decomposed into 1600 first-order
quadrilateral elements. We simulate the impact for 20~$\mu$s using an
adaptive timestep with the initial timestep set to 0.001~$\mu$s and
the maximum allowable timestep set to 0.012~$\mu$s for a total of
1,676 timesteps. The simulations are executed on the SGI ICE X high
performance computer "Topaz'' at the Engineering Research and
Development Center.

For simulations that use the sparse GP method, only a single compute
node, containing a 36 core 2.3 GHz Intel Xeon Haswell processor, is
used because all of the microscale model data is precomputed. The
microscale model is computed for values of $e$ between $-1.0 \times
10^7$~J/m$^3$ and $1.7 \times 10^8$~J/m$^3$ and for $\rho$ between
1.75~g/cm$^3$ and 1.93~g/cm$^3$. The bounds are selected based upon
previous experience running the simulation. The requirement to select
appropriate bounds for the grid sampling is a drawback of the sparse
GP method compared to the adaptive sampling method, which is able to
expand the sampling region during model evaluation on-demand. Two sets
of microscale model data are obtained corresponding to $q = 6$ and $q
= 7$, for a total of 4,225 and 16,641 points respectively. For each
level $q$, three Taylor impact simulations are performed under
different values of the sparsity parameter $R$: $R$ = 6, 8, and 10 for
$q=6$ and $R$ = 8, 10, and 12 for $q=7$.  The squared exponential
kernel
\[
k_{\sigma_n^2}(x, x^{\prime}) = \sigma_f^2 \exp\left(\frac{-|x - x^{\prime}|^2}{2l^2}\right) + \sigma_n^2 \delta(x, x^{\prime})
\]
is used for the GP, where the hyperparameter $l$ is the length-scale,
$\sigma_f^2$ is the signal variance and $\sigma_n^2$ is the noise
variance.  The choice of the squared exponential kernel reflects our
prior belief that the microscale model outputs are smooth and hence we
seek a smooth approximation.  Hyperparameters for the sparse GP
surrogate model are obtained by minimizing the negative log marginal
likelihood (NLML) with the BFGS algorithm implemented in the dlib C++
toolkit~\cite{dlib2009} using a stopping criterion of $10^{-3}$. The
starting point for the hyperparameter optimization is chosen to be
$l=0.2$, $\sigma_f^2=1.0$, and $\sigma^2_n = 0.01$.

Simulations employing the adaptive sampling method are executed on a
total of 90 compute nodes for a total of 3,240 cores. The macroscale
model and adaptive sampling module are executed on a single node, with
the remaining 89 compute nodes dedicated to microscale model
evaluations. The maximum number of points per local surrogate model,
$d_{max}$ is chosen to be $50$. Three simulations are performed under
different $\tilde{e}_{tol}$: $10^{-2}$, $5 \times 10^{-3}$, and
$2 \times 10^{-3}$.

The results of the adaptive sampling and sparse GP simulations are
compared to those of a reference simulation computing the
microscale DPD model for all EOS evaluations of the macroscale
model. The reference simulation requires a total of 2,681,600
microscale model evaluations and completes in 32.05~days of wall-clock
time on 12,852 processor cores for a total compute time of
9,885,192~hours.

\subsection{Results}

\begin{table}
\centering
\begin{tabular}[]{| c | c | c | c | c | c | c |}
\hline
\parbox[t]{.2cm}{\centering $q$} &
\parbox[t]{.2cm}{\centering $R$} &
\parbox[t]{1.05cm}{\centering Sparsity} &
\parbox[t]{1.65cm}{\centering Wall-Clock\\Time (hr)} &
\parbox[t]{2.15cm}{\centering Hyperparameter\\Opt. Time (hr)} &
\parbox[t]{1.4cm}{\centering Compute\\Time (hr)} &
\parbox[t]{2cm}{\centering \# Microscale\\Model Evals} \\
\hline
6 & 6 & 0.74 & 0.9 & 0.6 & 11,504 & 4,225 \\
\hline
6 & 8 & 0.64 & 1.5 & 0.8 & 11,525 & 4,225 \\
\hline
6 & 10 & 0.54 & 2.7 & 1.5 & 11,561 & 4,225 \\
\hline
7 & 8 & 0.85 & 11.4 & 8.8 & 44,345 & 16,641 \\
\hline
7 & 10 & 0.80 & 18.4 & 15.3 & 44,570 & 16,641 \\
\hline
7 & 12 & 0.74 & 40.9 & 32.3 & 45,274 & 16,641 \\
\hline
\end{tabular}
\caption{Timing data for simulations with sparse GP surrogate models. The sparsity is the fraction of non-zero entries in the original covariance matrix $K^{(q)}$ that are replaced by zeros in the sparse covariance matrix $K_R^{(q)}$. The wall-clock time is for the execution of the simulation and includes the optimization of the hyperparameters. The compute time includes both the time required to precompute the microscale model at sampling points and for the execution of the simulation.}
\label{table:sparsegp_results}
\end{table}

Timing data for Taylor impact simulations with the sparse GP surrogate
model are presented in Table~\ref{table:sparsegp_results}.  Here the
sparsity is the fraction of non-zeros in the covariance matrix
$K^{(q)}$ that are replaced by zeros in the sparse covariance matrix
$K_R^{(q)}$ (c.f. \eqref{eqn:sparse-K}).  The compute time includes
the time required to evaluate the microscale model for input points of
grid level $q$, the time to optimize the hyperparameters of the model,
and the time to execute the Taylor impact simulation. The compute time
to sample the microscale model at grid points is 11,475~hr for $q=6$
and 43,980~hr for $q=7$. This indicates that for all simulations with
the sparse GP surrogate model, the vast majority of total compute
time, greater than 95\%, is spent sampling the microscale model. We
note, however, that sampling the microscale model incurs a one-time
expense for a particular $q$ and the data can be reused across
multiple simulations. In addition, the sampling points are nested
across levels, which allows some microscale model data to be reused
from lower $q$ levels. Obtaining microscale model results at sampling
points is embarrassingly parallel as all points are chosen ahead of
time for a particular value of $q$. Given sufficient computing
resources, the microscale model data can be obtained in a very short
amount of wall-clock time. Therefore, the wall-clock time given in
Table~\ref{table:sparsegp_results} omits time spent precomputing the
microscale model and includes only the time to optimize
hyperparameters, dependent on the choice of $R$, and the time required
to complete the Taylor impact simulation. A significant portion of the
wall-clock time, ranging from 50\% to 83\% is spent on optimization of
the hyperparameters. The remaining time is spent on evaluation of the
surrogate model and the integration of the macroscale model forward in
time.

\begin{table}
\centering
\begin{tabular}[]{| c | c | c | c |}
\hline
\parbox[t]{.5cm}{\centering $\tilde{e}_{tol}$} &
\parbox[t]{1.65cm}{\centering Wall-Clock\\Time (hr)} &
\parbox[t]{1.4cm}{\centering Compute\\Time (hr)} &
\parbox[t]{2cm}{\centering \# Microscale\\Model Evals} \\
\hline
$10^{-2}$                & 2.3     & 7,546  & 833 \\
\hline
$5\times10^{-3}$         & 8.7     & 28,315 & 1,878 \\
\hline
$2 \times 10^{-3}$       & 156.3   & 506,333 & 27,827 \\
\hline
No Surrogate Module & 769.2   & 9,885,192 & 2,681,600 \\
\hline
\end{tabular}
\caption{Timing data for simulations with adaptive sampling. The wall-clock time is for the execution of the simulation, which includes the on-demand evaluation of the microscale model and update of the collection of surrogate models according to the adaptive sampling algorithm. The compute time is for the execution of the simulation, which for adaptive sampling includes the microscale model evaluation.}
\label{table:as_results}
\end{table}

For comparison, we present timing data in Table~\ref{table:as_results}
for simulations that use adaptive sampling. A detailed discussion of
these timings can be found in~\cite{leiter2018accelerated}. It should
be noted here that both the adaptive sampling method and the sparse GP
method allow for simulations that are orders of magnitude cheaper than
the benchmark simulation which always obtains the EOS from the
microscale model.

\begin{figure}
  \centering
  \includegraphics[width=0.5\textheight]{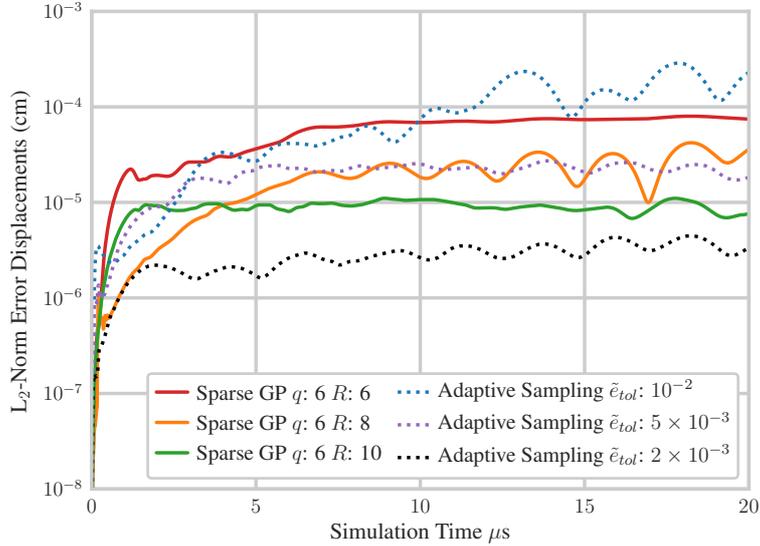}
  \caption{Error in displacements for three Taylor impact simulations
    with a sparse GP surrogate model with $q = 6$ and $R =$ 6, 8,
    and 10 and for three simulations using adaptive sampling with
    $\tilde{e}_{tol} = 10^{-2}, 5 \times 10^{-3},$ and $2 \times 10^{-3}$}
  \label{fig:level6_displacements}
\end{figure}

We now assess the effect of grid level $q$ and sparsity parameter $R$
on the solution of the Taylor impact problem. In
Figure~\ref{fig:level6_displacements}, we plot the $L_2$-norm of the
error in the displacements predicted by the macroscale model as a
function of simulation time for sparse GP simulations with $q=6$. For
comparison, we also plot the error for the adaptive sampling
simulations with $\tilde{e}_{tol}$ of $10^{-2}$, $5 \times 10^{-3}$
and $2 \times 10^{-3}$.  The $L_2$-norm of the error in the
displacements field $\vect{u}$ is:
\begin{equation}
|| \vect{u}^{surr} - \vect{u}^{ref} ||_2
   = \left(\int_V \sum_{i=1}^3 \lvert u_i^{surr}(v) -
   u_i^{ref}(v) \rvert^2 dv \right)^{1/2}
\end{equation}
where $u_i^{surr}(v)$ is the $i$-th component of the the
displacements field obtained from the simulation employing a surrogate model,
$u_i^{ref}(v)$ is the $i$-th component of the field from the
reference simulation, and $V$ denotes the volume of the cylinder.

As $R$ is increased, the error in displacements is reduced. The
sparse GP simulation with $q=6$ and $R=8$ has an error roughly
equivalent to the adaptive sampling simulation with $\tilde{e}_{tol} =
5 \times 10^{-3}$, but uses only 17\% of the wall-clock time and 41\%
the compute time. The simulation with $q=6$ and $R=10$ has a lower
error than the adaptive sampling simulation with $\tilde{e}_{tol}=5
\times 10^{-3}$, but requires only 31\% of the wall-clock time and
$41\%$ of the compute time. These results demonstrate that the sparse
GP method provides a more accurate solution than the adaptive sampling
method at a fraction of the wall-clock and simulation time.

For $q=6$, the number of microscale model evaluations is 4,225, which
is more than twice as many as used in the adaptive sampling simulation
with $\tilde{e}_{tol}=5 \times 10^{-3}$. It may appear surprising that
the cost of the sparse GP simulations, in terms of both wall-clock
time and compute time, are less than those using adaptive sampling
despite the many more microscale model evaluations used. However, the
reduction in cost is completely a consequence of the computational
load imbalance inherent in the adaptive sampling method that is not
present when using the sparse GP surrogate model. In adaptive
sampling, many processors are left underutilized during long stretches
of the simulation due to the unpredictable on-demand evaluation of
microscale models, a major drawback to the method. In the sparse GP
approach, all microscale model data is precomputed ahead of time and
is perfectly scalable and computationally efficient. Computational
resources are also fully utilized throughout the simulation itself,
spent primarily on the evaluation of the EOS using the surrogate
model.

\begin{figure}
  \centering
  \includegraphics[width=0.5\textheight]{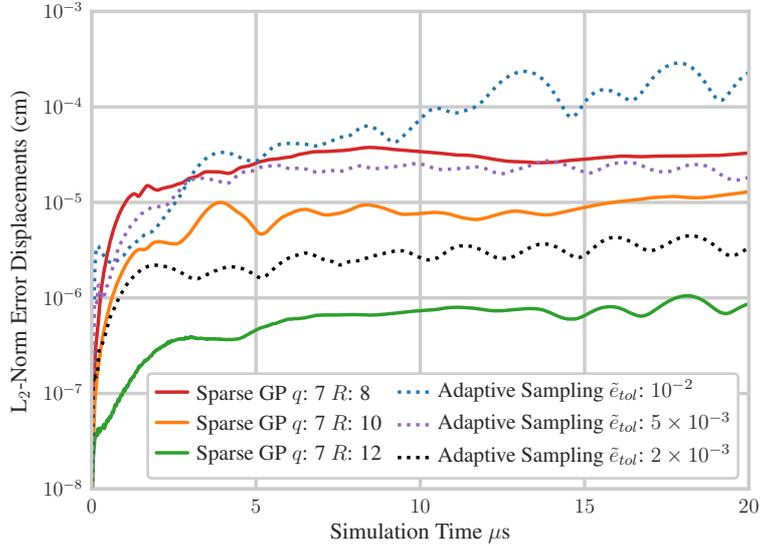}
  \caption{Error in displacements for three Taylor impact simulations
    with a sparse GP surrogate model with $q = 7$ and $R = 8, 10$,
    and 12 and for three simulations using adaptive sampling
    with $\tilde{e}_{tol} = 10^{-2}, 5 \times 10^{-3}$, and $2 \times 10^{-3}$}
  \label{fig:level7_displacements}
\end{figure}

None of the sparse GP simulations with $q=6$ reduce the error below
that obtained using adaptive sampling with the lowest
$\tilde{e}_{tol}=2 \times 10^{-3}$.  Further simulations with $q=6$
and values of $R$ beyond $10$, not included here, provide no further
reduction in the error. To determine whether the error of sparse GP
simulations can be reduced further by increasing the amount of
microscale model data available, the error in the displacements field
for sparse GP simulations with $q=7$ is plotted in
Figure~\ref{fig:level7_displacements}. In fact, the error is reduced
further by using a higher $q$. As was the case with $q=6$, increasing
$R$ reduces the error in the solution. The sparse GP simulation with
$q=7$ and $R=12$ provides an error in displacements well below the
smallest error obtained using adaptive sampling with
$\tilde{e}_{tol}=2 \times 10^{-3}$ and is able to achieve the reduced
error with 26\% of the wall-clock time and 9\% of the compute time, a
significant improvement. Furthermore, the sparse GP surrogate model
uses many fewer microscale model evaluations, only 16,641, compared to
the 27,827 microscale model evaluations required by the most accurate
adaptive sampling simulation. This is especially strong evidence of
the advantage of the sparse GP method. Given even fewer samples of the
microscale model, the sparse GP method produces a more accurate
surrogate model than adaptive sampling.

One interesting comparison can also be made between two of the sparse
GP simulations, one with parameters $q=6$ and $R=8$ and another with
$q=7$ and $R=8$. Both simulations have roughly the same error in the
solution, despite the increase in microscale model data available for
the $q=7$ simulation run. This observation can be explained by the
error bound \eqref{eqn:error-bounds}.  For sparse GP with $q = 7$, the
polynomial term $p(N)$ in the error bound may be significantly larger
than that of the sparse GP with $q = 6$, which offsets the higher
accuracy provided by the finer sampling.  This indicates that when
less accuracy is required, a good strategy may be to avoid over
sampling the microscale model as a coarse sampling should be
sufficient.

\section{Summary}
\label{sec:conclusion}
%%% Local Variables: ***
%%% mode:latex ***
%%% TeX-master: "acc_GP.tex"  ***
%%% End: ***

In this article, we have described a formulation of a sparse GP
regression method based on the near linear complexity sparse Cholesky
algorithm of Sch{\"a}fer~{\em et
  al.}~\cite{schafer2017compression}. The sparse GP method is capable
of utilizing a large number of training data and, thus, avoiding a
bottleneck associated traditionally with GP methods. This behavior is
due to the fact that the dense covariance matrix admits an approximate
sparse Cholesky decomposition, which reduces the computational
$\mathcal{O}(N^3)$ complexity down to near linear in $N$. In addition,
the sparse GP method provides error bounds for the sparse
approximation that are exponentially small with respect to a chosen
sparsity parameter.

Subsequently, we have employed the sparse GP method to construct a
surrogate model for a microscale model in a two-scale model of
deformation of an energetic material. The microscale model
characterizes the volumetric response of the energetic material by
recourse to dissipative particle dynamics. In turn, the macroscopic
model is a finite-element model simulating dynamic deformation. We
have considered two scenarios: 1) the evaluation of the microscale
model is replaced by an adaptive sampling approach, amounting to
on-the-fly construction of GP surrogates with compact support; 2) the
evaluation of the microscale model is replaced by a sparse GP
surrogate model constructed over the entire domain of the microscale
model. We have contrasted these two scenarios in terms of the overall
computational cost required to construct the surrogate models and
perform simulations with the macroscopic model.

Our results indicate that the sparse GP surrogate model offers
remarkable computational savings over the adaptive sampling surrogate
model. In some cases, these savings translate into over 10-fold
reduction in terms of the computational cost. This reduction is
primarily due to the fact that the sparse GP surrogate model relies on
a fixed grid for the selection of sampling points and the sampling can
be easily carried out with an extraordinary level of concurrency. Of
course, appropriate bounds for the extents of the grid must be
selected {\em a priori} as to guarantee that all evaluations requested
by the macroscopic model will ultimately be fully contained within the
grid. We emphasize that the adaptive sampling surrogate model does not
suffer from such a limitation as it is constructed adaptively and can
incorporate data from any subset of the domain. However, in practice,
good estimates for the bounds may be easily available from, for
example, lower fidelity microscale models.

In addition to the computational savings offered by the sparse GP
method, it also leads to a greater accuracy in the displacement field
than the local GP models constructed in the adaptive sampling
method. The increase in accuracy of the sparse GP method is likely
because it takes into account the long range correlations between data
points, which are not included when the data is partitioned into
separate GP surrogates with compact support.

It should be pointed out that the EOS regression problem showcased
here is low dimensional and that additional methods will be required
to extend GP regression to high dimensions, such as active
subspaces~\cite{tripathy2016Gaussian} or additive
kernels~\cite{duvenaud2011Additive}. Sparse GP regression helps to
alleviate the problem of high dimensionality somewhat by enabling the
incorporation of more data into the surrogate model. Perhaps combining
dimension reduction methods with the sparse GP method outlined
here could address even higher dimensional problems and is a subject
of ongoing research.

%% It should be understood however, that GP regression will be an
%% unlikely candidate for constructing surrogate models in high
%% dimensions. Moreover,  new approaches will need to be
%% explored. \textcolor{red}{Anything else to add?}

%%% Local Variables:
%%% mode: latex
%%% TeX-master: "acc_GP.tex"
%%% End:

\section{Acknowledgements}
\label{sec:acknowledgements}
The authors would like to thank Dr. Rich Becker of the U.S. Army
Research Laboratory and Florian Sch{\"a}fer of Caltech for fruitful
discussions. The work of T.W. and P.P. was supported in part by the
DARPA project W911NF-15-2-0122. This work was supported in part by a
grant of computer time from the DoD High Performance Computing
Modernization Program at the U.S. Army Engineer Research and
Development Center.

\section*{References}
\bibliography{mybibfile}

\end{document}